\documentclass{article}
\title{{\bf On Schr\"odinger's cat}
\vspace*{-1cm}
\begin{tabular}{c}
\normalsize  Nalin de Silva \\
\normalsize   Department of Mathematics, University of Kelaniya, Kelaniya, Sri Lanka\\
\end{tabular}
\date{}
}
\usepackage{amsmath,amssymb}
\begin{document}
\maketitle
\section*{Abstract}
Schr\"odinger's cat appears to have been harassed in a chamber during the past eighty years or so by interpreting the role of the observer as a person, who sets an experiment and then observes results, may be after some time.  The realist position tells us that the physical processes would take place independent of the observer with well defined properties, whereas the positivist position wants us to believe that nothing can be said of a system when it is not being observed. In this paper we question both these positions and also the assumption that the atom and the cat are entangled and further whether the atom could be considered to be in a state of decay and not decay. We let the cat either out of the bag (chamber) or rest in peace with or without the atom or the observer.

\section{INTRODUCTION}

In the thought experiment designed by Schrödinger a radio active element is kept in a tube and a mechanism is set to free a poisonous gas when an atom decays killing the cat. The experiment is described in many books and the reader is referred to Baggott$^1$  for a good account of it. It is assumed that when the system is kept in a closed chamber the atom is in a state of superposition of decayed and undecayed states and hence the cat is in a state of superposition of living state and dead sate. Schrödinger argued that according to the Copenhagen Interpretation as long as the observer does not look at the system the atom is both decayed and undecayed and the cat that is entangled with the atom is both alive and dead. Only an observation (measurement) by the observer would reveal whether the cat is dead or alive. To Schrödinger this conclusion was not acceptable and he and especially Einstein found the Copenhagen Interpretation to be incomplete.

\section{ENTANGLEMENT OF THE CAT AND THE ATOM}

There are some important assumptions made by Schrödinger in describing the thought experiment, which should draw our attention. The first and foremost is the entanglement of the atom and the cat. It is generally believed that Schrodinger coined the word entanglement for the first time in describing this particular thought experiment.  However, de Silva$^2$  has argued that the cat and the atom are not entangled. If there was no poisonous gas there was no possibility for the cat to die as a result of the decay of the atom. On the other hand the cat could die even if the atom was not decayed due to other reasons. Even if the chamber is filled with air and there is sufficient oxygen for the cat to survive, it could have had a ``normal'' death even before the atom decayed. Even if one were to ignore arguments based on micro and macro nature of the atom and the cat respectively and whether Quantum Mechanics could be applied to macro objects, one could not ignore the fact that the cat and the atom are not entangled, contrary to what Schrodinger had assumed. If the cat had died before an atom decayed, would it be still in a state of superposition of death and alive before the observer opens the chamber? As long as the cat also has an existence or death independent of the atom it is not possible to conclude that the cat and the atom are entangled. We preserve the term entanglement to refer to two or more systems or particles that are connected to each other in such a way that changing a property of one of them correspondingly affects the others simultaneously no matter what the distance between them may be.  The systems or particles may made to be entangled but once they are entangled they cease to have an existence ``independent'' of each other. The atom and the cat under discussion have existences independent of each other. 

Suppose that when the chamber is opened the observer finds the cat to be dead. There are two possibilities as far as the atom is concerned. It could be either decayed or not. If the atom is not decayed the observer could come to the conclusion that the cat had died not as a result of the decay of the atom but due to ``natural causes''. However, if the atom is decayed there are two possibilities and the cat could have died due to ``natural causes'' or as a result of the decay of the atom. A post-mortem investigation has to be carried out in order to decide the cause, and one could even say that until this particular investigation is made the cat could be supposed to have died both due to ``natural causes'' and as a result of the decay of the atom. However, that situation is possible only if the cat dies simultaneously due to both, as a dead cat cannot die a second time even if has nine lives.

\section{OBSERVATION AND COLLAPSE OF THE WAVE FUNCTION}

The above leads to a discussion on observation of Quantum Systems and collapse of the wave function in Quantum Mechanics. What is meant by the statement that a system is in a state of superposition? Does the wave function of the system collapse only when a conscious observer is able to observe it? Can it collapse without conscious observers observing the system? What was the situation in the world (universe) before conscious observers came into existence, assuming that they were not there from the beginning of the world (universe)? Even if the conscious observers have been present from the very beginning, did quantum systems remain in multiple superposition states before the conscious observers were able to observe them? Is it possible even today to have some quantum systems in the universe that have not been observed by conscious observers in multiple superposition states? 

It appears that the relationship between the observed and the observer has not been clearly understood in Quantum Mechanics. Chandana and de Silva$^3$  in their new interpretation of Quantum Mechanics make a distinction between existence of a Quantum Mechanical system and its observation. It is suggested in this paper that it is not necessary to have a conscious observer for the wave function of a quantum mechanical system to collapse. The wave function could collapse even without a conscious observer and the quantum mechanical system could be in one particular state and not in a multiple superposition state after an interaction with another particle or a system, without the observer being aware of what has transpired inside the chamber. Thus the cat even if it is ``entangled'' with the atom is either dead or alive depending on whether the atom has decayed or not. The atom is also in one of the states of either decay or non decay though the observer may have no knowledge of the situation. The system could collapse without a conscious observer in the sense that the observer does not have a knowledge of the collapse until he decides to observe the system. Here a distinction is made between the collapse of the wave function and the knowledge of the collapse by the observer.  

One would say that this situation arises even classically and would wonder as to the difference between Classical Mechanics and Quantum Mechanics. For example classical particles could interact with each other in the absence of a conscious observer without the observer having any knowledge of the interactions. Thus it appears that there is no difference between Classical Mechanics and Quantum Mechanics and the role of the observer is not significant in Quantum Mechanics. What is suggested here is that though the observer in Quantum Mechanics does not play a role different from an observer in Classical Mechanics there is a difference between Classical and Quantum states. The difference is that in Classical Mechanics a system is always in one definite state with respect to a property but in Quantum Mechanics the system could be in a multiple superposition sate with respect to a given property such as momentum or position at a given time. The next moment the system could collapse and be in one definite state with respect to the property that is considered but would be in a multiple superposition state with respect to the conjugate property. It could also be in a multiple superposition state with respect to some other property (or properties).    

The collapse of the wave function has nothing to do with the conscious observer. If it was the case, as mentioned before, the whole world (universe) would have been in multiple superposition states with respect to quantum properties, before conscious observers came into existence. It also means that when humans look at distant galaxies which had not been observed by any conscious observers the wave functions representing quantum systems collapse for the first time since the creation of the systems. This is not acceptable and it is clear that the role of the conscious observer has been misunderstood due to the influence of positivism. 

Albert$^4$    has stated the problem we face very clearly in the following manner. ``The dynamics and the postulate of collapse are flatly in contradiction with one another ... the postulate of collapse seems to be right about what happens when we make measurements, and the dynamics seems to be bizarrely wrong about what happens when we make measurements, and yet the dynamics seems to be right about what happens whenever we aren't making measurements.'' It is true that measurements are made by conscious observers but do quantum mechanical systems interact with other such systems or classical systems in the absence of such observers? The positivistic response to this question would be ``it does not matter whether such interactions take place or not in our absence as we are not aware of them''. However the logical positivists would dissociate from this view and we have to make a clear distinction between our knowledge of a property of a quantum mechanical system and its existence. It has to be emphasized that the positivistic attitude is not limited to quantum systems. One could easily say that there was no moon before any conscious observer was able to observe it in particular, or that there were no classical systems or particles before there were any conscious observers in general.   

However, it is only as far as quantum systems are concerned that this question is being debated in the world of Physics, though philosophers have not made up their minds in general in either the classical or the quantum world. This is due to the existence of multiple superposition states in quantum mechanics. When a system evolves in a multiple superposition state with respect to some property but is observed only in one specific state naturally the question arises as to when this change occurs. Bohr and Heisenberg, especially the latter, were of the view that it is at the point of observation that this change occurs. Initially Bohr may not have thought the same way as Heisenberg but it appears that gradually he had come to embrace the position of Heisenberg. It is clear that Heisenberg has been influenced by positivism and would not have wanted to speculate on what goes in before an observer makes an observation or takes a measurement. However, most of the Physicists were not prepared to subscribe to Heisenberg's views though they saw a problem in the collapse of the wave function. 

It was obvious that the instrument that interacts with the quantum system had to play a role here but instead of assuming that the collapse of the wave function takes place at the instance of interaction of the system and the instrument they thought that it would happen after some time and were prepared to suggest elaborate schemes (theories) to understand this problem. Von Neumann was the first to analyze this problem mathematically. In what follows a brief description of the method as given by Baggott$^1$ is stated in order to illustrate the reasoning behind this so called entanglement of the quantum mechanical system and the instrument. (The numbering of equations and references has been left as it appears in the original.)

``Suppose a quantum system described by some state vector $|\mathbf{\Psi}>$  interacts with a measuring instrument whose measurement eigenstates are $|\psi_+>$  and  $|\psi_->$ . These eigenstates combine with the macroscopic instrument to reveal one or other of the two possible outcomes, which we can imagine to involve the deflection of a pointer either to the left ($+$ result) or the right ($-$ result). Recognizing that the instrument itself consists of quantum particles, we describe the state of the instrument before the measurement in terms of a state vector  $|\phi_0>$, corresponding to the central pointer position. The total state of the quantum system plus the measuring instrument before the measurement is made is described by the state vector  $|\Phi_0>$, which is given by the product:
\begin{eqnarray*}
|\Phi_0>=|\Psi>|\phi_0>&=& \frac{1}{\sqrt{2}}[|\psi_+>+|\psi_->]|\phi_0>\nonumber\\
&=& \frac{1}{\sqrt{2}} [|\psi_+>|\phi_0>+|\psi_->|\phi_0>] \text{\qquad \qquad  \qquad   (3.2)}
\end{eqnarray*}
where we have made use of the expansion theorem to express $|\mathbf{\Psi}>$   in terms of the measurement eigenstates and we have assumed that  $<\psi_+|\mathbf{\Psi}>=<\psi_-|\mathbf{\Psi}>=1/\sqrt{2}$  (the results are equally probable).

	We want to know how  $|\mathbf{\Phi}_0>$ evolves in time during the act of measurement. From our discussion in Section 2.6, we know that the application of the time evolution operator  $\hat{U}$ to  $|\mathbf{\Phi}_0>$ allows us to calculate the state vector at some later time, which we denote as  $|\mathbf{\Phi}>$, according to the simple expression  $|\mathbf{\Phi}>=\hat{U}|\mathbf{\Phi}_0>$, or
\begin{equation*}
|\mathbf{\Phi}>=\frac{1}{\sqrt{2}}[\hat{U}|\psi_+>|\phi_0>+\hat{U}|\psi_->|\phi_0>] \tag{3.3}
\end{equation*}
We now have to figure out what the effect of  $\hat{U}$ will be.

	It is clear that if the instrument interacts with a quantum system which is already present in one of the measurement eigenstates ($|\psi_+>$, say), then the total system (quantum system plus instrument) must evolve into a product quantum state given by  $|\psi_+>|\phi_+>$. This is equivalent to saying that this interaction will always produce a $+$ result (the pointer always moves to the left). In this case, the effect of $\hat{U}$  on the initial product quantum state  $|\psi_+>|\phi_0>$  {\it must} be to yield the result  $|\psi_+>|\phi_+>$, 
i.e.
\begin{equation*}
\hat{U}|\psi_+>|\phi_0>=|\psi_+>|\phi_+>. \tag{3.4}
\end{equation*}
Similarly,
\begin{equation*}
\hat{U}|\psi_->|\phi_0>=|\psi_->|\phi_->.\tag{3.5}
\end{equation*}
Substituting these last two expressions into eqn (3.3) gives
$$|\mathbf{\Phi}>=\frac{1}{\sqrt{2}}[|\psi_+>|\phi_+>+|\psi_->|\phi_->] \text{ \qquad    (3.6)"}$$
\vspace*{.5cm}

It is clear that the problem is not solved by proceeding along these lines. It does not give the time at which the wave function collapses and there are at least two more fundamental objections that could be raised against this scheme. The measuring instrument is never in a multiple superposition state and it gives only one reading at any given time. The evolution of the wave function that describes the instrument cannot be described in terms of superposition states and though the instrument is made of quantum particles, it does behave as a classical object without multiple superposition states. The second objection is the absence of the observer in the scheme though implicitly assuming that the observer plays an important role in the collapse of the wave function. It is clear that the Physicists, though not quite agreeing with Copenhagen Interpretation have not been able to overcome the influence exerted by Heisenberg and his positivism. The wave function though considered as the tensor product of two state vectors is implicitly the tensor product of three state vectors including that of the observer. The never ending process associated with the above scheme comes to an end, as assumed implicitly, at the time of observation. 

There appears to be only one solution to the problem. The collapse of the wave function takes place either as a result of an external influence at the moment of interaction of the quantum mechanical system with an external object, which could be the instrument or any other external object or merely as a result of an internal ``mechanism''.  In the first instance, an observer is not necessarily needed as a quantum mechanical system could interact with another quantum or classical system without the assistance of a conscious observer. On the other hand an observer may set up the instrument and go away without getting any readings. The quantum mechanical system and the instrument do not have to wait for the observer to return to the scene to observe to produce a reading. 

When a quantum mechanical system interacts with an external body (instrument or something else) only two things could happen. Either the system changes as a result or not. It cannot change and not change at the same time unless the system can be in a state of superposition with respect to change and not change. However, it amounts to a combination of collapse of the wave function and not collapse of the same which appears to be at a higher level of superposition not theorized in quantum mechanics. If the quantum mechanical system changes, then the wave function collapses with respect to a given property and that particular property takes one value as a result. This is decoherence and many experiments have been done in the recent past on this aspect. The state function that evolves coherently (unitarily) with respect to a certain property interacts with an external system (classical or quantum, observer set up or not) and the evolution is changed irreversibly at the time of interaction. The observer may not have a knowledge of the change but it does not imply that the change has not taken place. The observer could find out for himself what has taken place afterwards by taking readings or observing through other means.   
It has to be emphasised that a quantum mechanical system could also change without any external influence including an observer as in the case of decay of atoms and these as well as changes due to external influences are governed by Born's probabilistic rule involving inner products of the state vectors. 

\section{DISCUSSION}
Does the above description mean that the observer has no role to play in quantum mechanics? Does it mean that the realist position is held by the above description? The answers to both the questions are not in the affirmative as has been shown by de Silva$^5$. The realist position is not tenable as local reality is not adhered to following experiments of Aspect$^6$    and many others since then following the celebrated work by Bell$^7$. It is not possible to predict the results of a measurement before hand and unlike in classical systems one has to resort to probabilities in quantum systems. Before an interaction the quantum mechanical system may be in a multiple superposition state that cannot be observed but whose ``existence can be known'' by the observer as has been shown by Monroe et al$^8$  and by many others subsequently. If the observer wants to find exactly the particular state then the system has to be disturbed using external means and decoherence sets in.

This does not mean that the observer has no role to play at all. We live in a conceptual world and not in a so called objective realist world. All our observations depend on our concepts and in turn concepts are created in order to explain the world we observe. Observations and creations of concepts by observers are interconnected and are not independent of each other. The observer is involved not only in the case of  quantum mechanical systems but with respect to classical mechanical systems as well, in the sense that concepts such as momentum, position are all his creations and do not have an objective existence as such. This has been discussed in more detail by de Silva$^5$, and the outlook in that paper as well as in the present paper is neither realist nor positivist.

\vspace*{5mm}
REFERENCES\\
\hrule
\begin{enumerate}
\item[1.]    Baggott J., The meaning of Quantum Theory, Oxford University Press, 1992
\item[2.] de Silva  Nalin, What's wrong with Schr\"odinger's cat, Proceedings of the Annual Research Symposium, University of Kelaniya, 2003, 60    
\item[3.] Chandana S., and de Silva Nalin, A new interpretation of Quantum Mechanics, Proceedings of the Annual Research Symposium, University of Kelaniya, 2004, 59
\item[4.]Albert D., Quantum Mechanics and Experience, Cambridge, MA: Harvard university Press, 1992
\item[5.]de Silva Nalin, Quantum Physics in a different ontology, arXiv:1006.4712 , 2010
\item[6.]Aspect A., Physical Review Letters, 1982, 49
\item[7.]Bell, J., Physics (N Y) 1, 1965, 195
\item[8.]Monroe C. et al., A ``Schrödinger Cat'' superposition state of an atom, Science, 1996, 272
\end{enumerate}                                                                                                                   
   
\end{document}